\documentclass[a4paper,10pt]{article} 

\usepackage[left=2.3cm,top=2.5cm,right=2.3cm,bottom=0.8in]{geometry} 
\usepackage{bold-extra}
\usepackage{color}
\usepackage{graphicx}
\usepackage{wrapfig}
\usepackage{xspace}
\usepackage{array}
\usepackage[english]{babel}
\usepackage{titlesec}
\usepackage{xcolor,colortbl}
\usepackage{hhline}
\usepackage{makecell}
\usepackage[bookmarks=true,pdfborder={0 0 0}]{hyperref}
\usepackage{amsmath}
\usepackage{amsfonts}
\usepackage{amssymb}
\usepackage{esint}
\usepackage{cite}
\usepackage[sort&compress,square,comma,authoryear]{natbib}
\definecolor{myblue}{rgb}{0,0,0.8} 
\newcommand{\be}{\begin{equation}}
\newcommand{\ee}{\end{equation}}
\newcommand{\bea}{\begin{eqnarray}}
\newcommand{\eea}{\end{eqnarray}}
\newcommand{\bes}{\begin{equation}\begin{split}}
\newcommand{\ees}{{\end{split}\end{equation}}}

\renewcommand{\S}{Section~}

\newcommand{\msun}{{\rm M}_{\odot}}

\newcommand{\ha}{H{\sc\,i}\xspace}
\newcommand{\hm}{H$_2$}

\newcommand{\Ms}{M_{\ast}}

\newcommand{\Mha}{M_{\rm H{\sc\,I}}}
\newcommand{\Mhm}{M_{\rm H_2}}

\newcommand{\sax}{S$^3$-SAX\xspace}

\newcommand{\mnras}{MNRAS} 
\newcommand{\nat}{Nat} 
\newcommand{\apj}{ApJ} 
\newcommand{\pasp}{PSPS} 

\newcolumntype{L}[1]{>{\raggedright\let\newline\\\arraybackslash\hspace{0pt}}m{#1}}
\newcolumntype{C}[1]{>{\centering\let\newline\\\arraybackslash\hspace{0pt}}m{#1}}
\newcolumntype{R}[1]{>{\raggedleft\let\newline\\\arraybackslash\hspace{0pt}}m{#1}}
\title{\vspace{-14mm}100 deg$^2$ Mock Galaxy Cone for \ha Surveys with the Early SKA\\[-1ex]}
\date{\vspace{-1ex}\today}
\author{D.~Obreschkow and M.~Meyer\\[0ex]International Centre for Radio Astronomy Research (ICRAR)}

\begin{document}
\pagenumbering{gobble} 

\maketitle

\vspace{-4mm}

This document accompanies an easy-to-use mock catalog of galaxies with detailed neutral \emph{atomic} hydrogen (\ha) and auxiliary \emph{molecular} and \emph{optical} properties. The catalog covers a field of 10-by-10 degrees and a redshift range of $z=0-1.2$. It contains galaxies with 21\,cm peak flux densities down to $1\,\mu\rm Jy$ and is, within this flux limit, complete for \ha masses above $10^8\msun$. Five random realisations of the catalog in ASCII format ($\sim$4\,GB/file) and subtables with \ha flux limits of $10\,\mu\rm Jy$ ($\sim$500\,MB/file) and $100\,\mu\rm Jy$ ($\sim$30\,MB/file) can be downloaded at\vspace{1.5mm}

{\centering\href{http://ict.icrar.org/store/staff/do/s3sax}{\textcolor{myblue}{\Large http://ict.icrar.org/store/staff/do/s3sax}}\\}
\vspace{5.5mm}
\emph{Context:} This catalog was released in the frame of the 2014 renewal of the science case for the Square Kilometre Array (SKA) telescope. It is the basis for performance calculations in several chapters of this science case and can be considered a reference tool to assess future surveys with the SKA and its pathfinders.

\emph{Model:} The catalog is a light version of the Semi-Analytic Suite of the SKA Simulated Skies (\sax), delivered as part of the European SKA design studies (SKADS) in 2009. It relies on the physical models described in \cite{Obreschkow2009b,Obreschkow2009e,Obreschkow2009f}, applied to the \emph{evolving} semi-analytic model by \cite{DeLucia2007} on the Millennium dark matter simulation \citep{Springel2005}. More sophisticated models and models run with updated cosmological parameters have since been produced, but the differences to \sax are minor in the considered redshift range. The \sax model has the advantage of having been exhaustively tested in more than a hundred peer-reviewed publications, successfully reproducing \ha masses, line profiles, Tully-Fisher relation, disk sizes, angular momentum, clustering and evolution characteristics. 

\emph{Cosmic volume:} The volume of the catalog is a cone with a 10-by-10 degree opening, truncated at a redshift of $z=1.2$. For \ha, this $z$ limit matches the 650~MHz lower limit of SKA1-Survey band~2 specified in the System Baseline Design (March 2013). Five independent cones are available to assess the cosmic variance within this volume. They are centred on the coordinates $\rm RA=36/108/180/252/324^\circ$ and $\rm Dec=-30^\circ$.  

\emph{Flux/mass limit:} The catalog has been truncated to 21\,cm peak flux densities above $1\,\mu\rm Jy$ to keep the catalog size reasonable. Within this flux limit, the model is complete for \ha masses above $10^8\msun$. At smaller \ha masses, the number of galaxies drops steeply, missing most of the \ha in dwarf galaxies. This limit is only a minor concern when dealing with isolated direct \ha detections in blind surveys at $z>0$, where only a tiny fraction of the total survey volume is sensitive to \ha masses $<10^8\msun$. However, when dealing with global \ha mass estimates, e.g. in a stacking experiment to measure $\Omega_{\rm H{\sc\,I}}(z)$, the \ha mass contained in unresolved galaxies is almost certainly significant (cf.\ \S3.4 of \citealp{Obreschkow2011c}), as apparent from direct comparisons of the modelled $\Omega_{\rm H{\sc\,I}}(z>1)$ against inferences from DLA data, hydrodynamical simulations by \cite{Pontzen2008} and statistical extrapolations by \cite{Lagos2011}. It is recommended that applications relying on the \emph{total} \ha content of the universe at $z\sim1$ correct for the \ha mass resolution limit of the catalog -- most simplistically by scaling all \ha masses in the mock catalog with a $z$-dependent factor ($>1$) to match $\Omega_{\rm H{\sc\,I}}(z)$ inferred from the best DLA and/or stacking data.

\emph{Optical properties:} Whereas the original \sax release did not contain optical properties, this new catalog includes approximate optical radii, stellar masses and apparent Vega $R$-band magnitudes, derived from simulated extinction corrected absolute magnitudes using a variable $K$-correction, fitted to empirically determined $K$-corrections as a function of redshift and morphology \citep{Driver1994,Westra2010}. In a test sample of $10^3$ mock galaxies, these apparent magnitudes were consistent with the `exact' magnitudes obtained by redshifting empirical SEDs fitted to the simulated absolute Vega BVRIK magnitudes.

\emph{Need additional properties?} This catalog has intentionally been limited to 20 properties per galaxy. Additional properties listed on \href{http://s-cubed.physics.ox.ac.uk/s3\_sax/sky}{http://s-cubed.physics.ox.ac.uk/s3\_sax/sky} (e.g.\ alternative CO line properties, surface density profiles, metallicity, black hole properties) can be downloaded there, respectively requested directly from the authors of this document.

\emph{We acknowledge Simon Driver for his hands-on assistance in calculating the $K$-corrections.} 

\newpage
\noindent\begin{tabular}{C{10mm}C{13mm}C{15mm}C{2mm}L{124mm}}
\hline\hline
Col & Symbol & Unit && Description \\
\hline
1 & ID & -- && Unique galaxy identifier in the Munich Semi-Analytic Model ``DeLucia2006a''\\
2 & RA & deg && Right ascension of galaxy centre \\
3 & Dec & deg && Declination of galaxy centre \\
4 & $z$ & -- && Apparent redshift of galaxy centre, including the Doppler component due to peculiar motion relative to the Hubble expansion \\
5 & $i$ & deg && Galaxy inclination defined as the smaller angle ($0^\circ-90^\circ$) between the line-of-sight and the rotational axis of the galaxy \\
6 & $T$ & -- && Numerical Hubble type ($-6...0$ for ellipticals, $0...10$ for spirals, $99$ for morphologically unresolved objects, mostly dwarfs) \\
7 & $\Ms$ & $\msun$ && Stellar mass \\
8 & $\Mha$ & $\msun$ && Mass of neutral atomic hydrogen \ha, without helium \\
9 & $\Mhm$ & $\msun$ && Mass of molecular hydrogen \hm, without helium \\
10 & $S_{\rm H{\sc\,I}}^{\rm int}$ & Jy\,km\,s$^{-1}$ && Velocity-integrated flux of the redshifted 21\,cm \ha emission line, with velocity units defined in the galaxy rest-frame \\
11 & $S_{\rm H{\sc\,I}}^{\rm peak}$ & Jy && Peak flux density of the \ha emission line; typically the flux density of the `horns' \\
12 & $S_{\rm CO}^{\rm int}$ & Jy\,km\,s$^{-1}$ && Velocity-integrated flux of the redshifted 115.27\,GHz $^{12}$CO(1--0) emission line, with velocity units defined in the galaxy rest-frame \\
13 & $S_{\rm CO}^{\rm peak}$ & Jy && Peak flux density of the $^{12}$CO(1--0) emission line; typically the flux density of the `horns' \\
14 & $W_{\rm H{\sc\,I}}^{50}$ & km\,s$^{-1}$ && Width of the \ha emission line, in galaxy rest-frame velocity units, measured at 50\% of the peak flux density\\
15 & $W_{\rm H{\sc\,I}}^{20}$ & km\,s$^{-1}$ && Width of the \ha emission line, in galaxy rest-frame velocity units, measured at 20\% of the peak flux density\\
16 & $r_{\rm H{\sc\,I}}^{\rm edge}$ & arcsec && Apparent \ha radius along the major axis out to a \ha disk surface density of $1\,\msun\rm pc^{-2}$, corresponding to a face-on column density of $1.25\cdot10^{20}\rm cm^{-2}$\\
17 & $r_{\rm H{\sc\,I}}^{\rm half}$ & arcsec && Apparent \ha half-mass radius along the major axis\\
18 & $M_{\rm R}$ & mag && Absolute Vega $R$-band magnitude, corrected for intrinsic dust extinction; 99 if stellar mass and star formation history are insufficiently resolved to compute $M_{\rm R}$\\
19 & $m_{\rm R}$ & mag && Apparent Vega $R$-band magnitude; value 99 if no absolute magnitudes available \\
20 & $r_{\rm e}$ & arcsec && Effective radius, here approximated as the radius containing half the stellar mass if the galaxy were viewed face-on\\
\hline\hline
\end{tabular}\\

\vspace{-2mm}
{\centering Table 1: Description of the columns of mock catalog in ASCII format.\\}

\vfill

\newpage

\begin{figure}[t]
	\centering
	\vspace{-7mm}
	~~~~~~\includegraphics[width=0.98\textwidth]{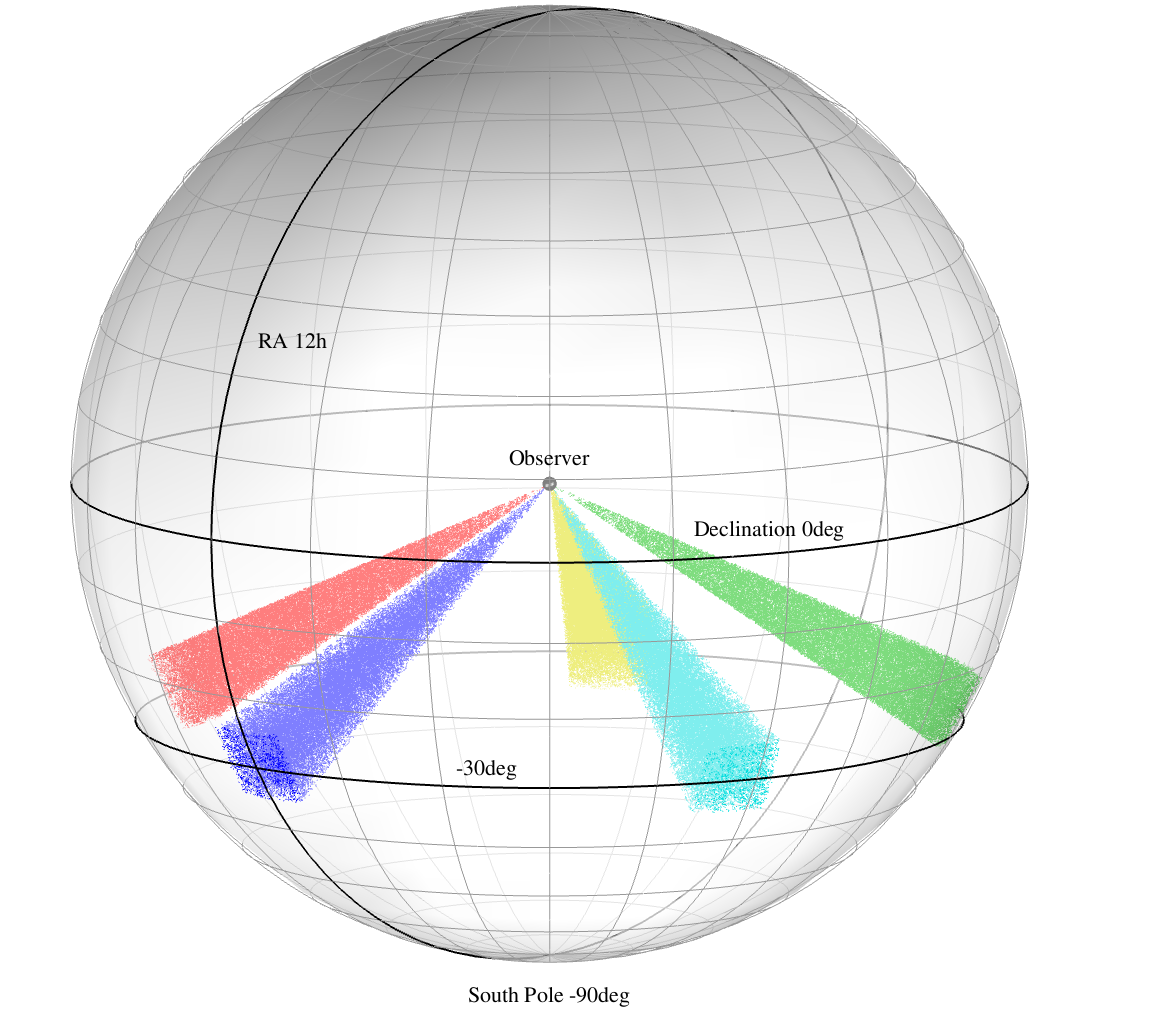}
	\vspace{-7mm}
	\caption{\small Representation of the five statistically independent mock cones of galaxies. The cones are centred on the coordinates $\rm RA=36/108/180/252/324^\circ$ and $\rm Dec=-30^\circ$.}
	\label{fig_sphere}
\end{figure}

\begin{figure*}[t]
	\centering
	\begin{tabular}{lC{4mm}r}
		\includegraphics[width=8.05cm]{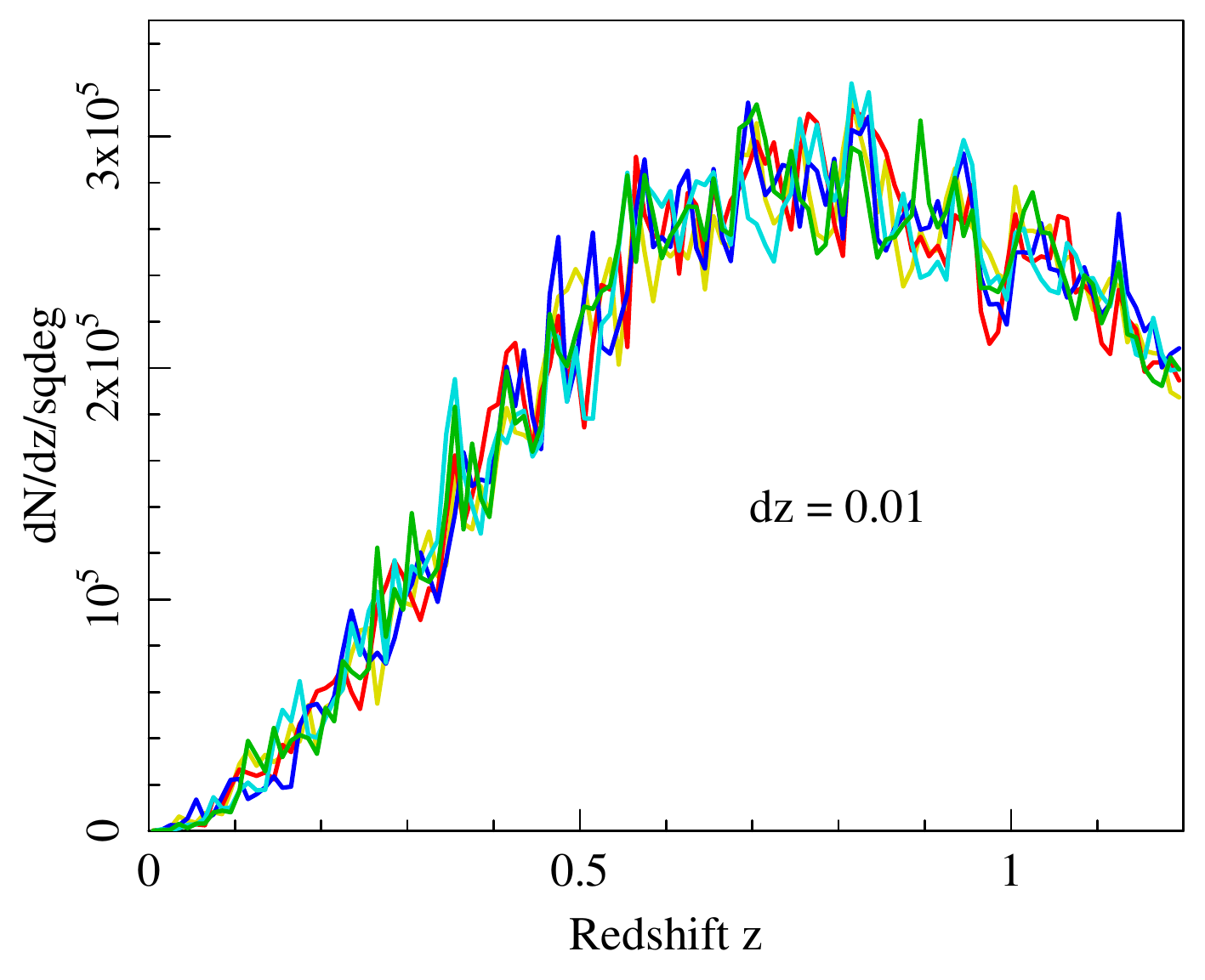} && \includegraphics[width=8.05cm]{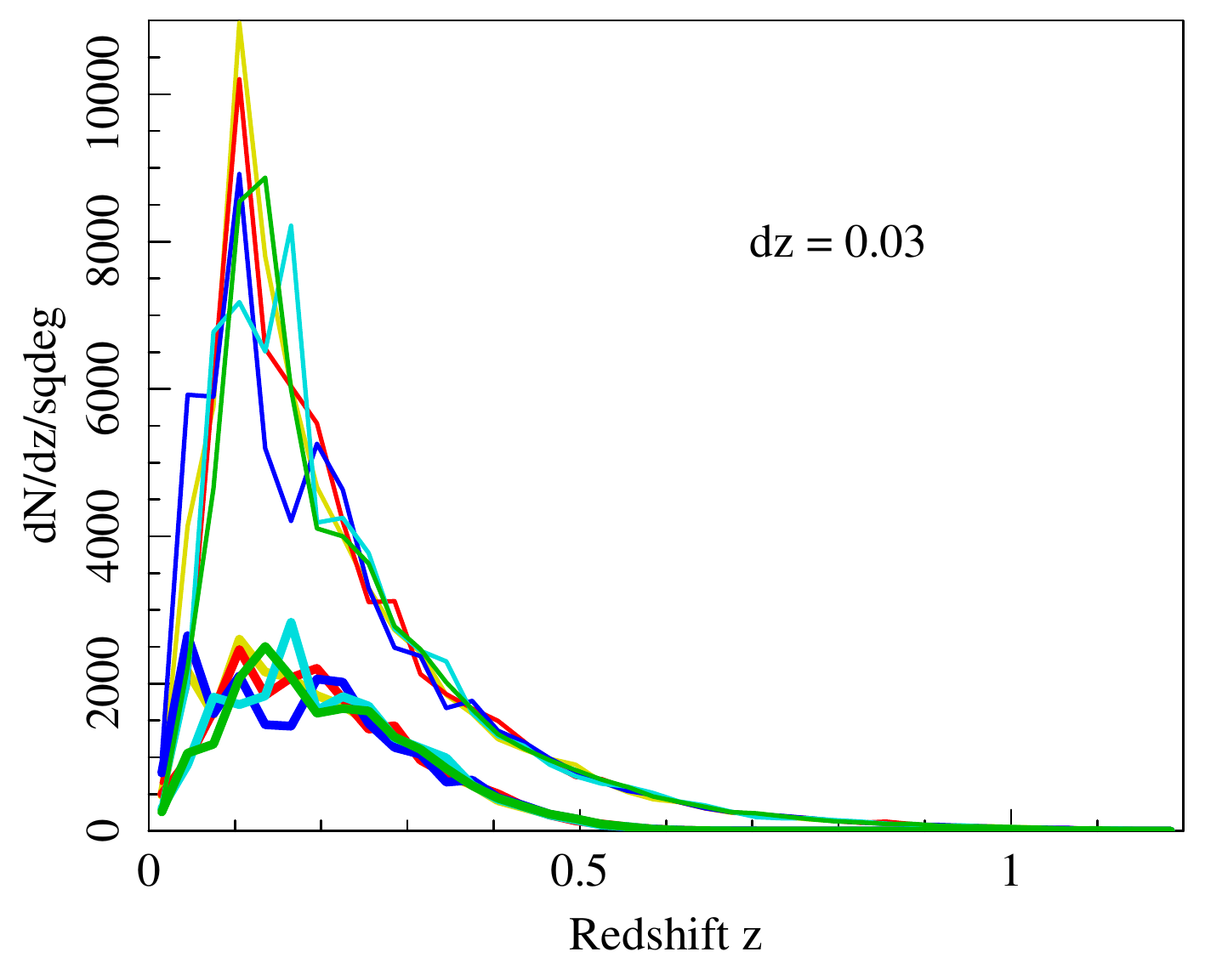}
	\end{tabular}
	\vspace{-5mm}
	\caption{\small Redshift distribution of galaxies, where the y-axis shows $dN/dz$ per \emph{square degree}, averaged from the 100 square degree field of view. The five colours represent the five cones shown in Fig.~\ref{fig_sphere}. LEFT: all galaxies in the full ($S_{\rm H{\sc\,I}}^{\rm peak}\geq1\,\rm\mu Jy$) cones. RIGHT: all galaxies with $S_{\rm H{\sc\,I}}^{\rm peak}\geq100\,\rm\mu Jy$ (thin lines) and subsamples of these galaxies (thick lines) with an apparent $R$-band magnitude $m_{\rm R}\leq20$ and an effective radius $r_{\rm e}\geq1''$, as an example of matching \ha data to a GAMA-like optical survey to derive approximate stellar mass and inclination measurements.}
	\label{fig_fwhm}
\end{figure*}

\end{document}